\documentclass[12pt]{article}
\usepackage{a4wide}
\usepackage{epsf}
\usepackage{rotating}

\def\r{\rho}

\def\H{{\cal L}}

\def\ra{\rightarrow}

\begin{document}

\title{Entanglement Entropy and The Density Matrix
Renormalization Group}

\author{Jos\'e Gaite\\
{\small\em Instituto de Matem{\'a}ticas y F{\'\i}sica
Fundamental,}
{\small\em CSIC, Serrano 123, 28006 Madrid, Spain}
}
\maketitle

\begin{abstract}
Quantum entanglement entropy has a geometric character. This is
illustrated by the interpretation of Rindler space or black hole
entropy as entanglement entropy.  In general, one can define a
``geometric entropy", associated with an event horizon as a boundary
that concentrates a large number of quantum states.  This allows one to
connect with the ``density matrix renormalization group" and to unveil
its connection with the theory of quantum information.  This
renormalization group has been introduced in condensed matter physics
in a heuristic manner, but it can be conceived as a method of
compression of quantum information in the presence of a horizon.  We
propose generalizations to problems of interest in cosmology.
\end{abstract}
     
\section{Entanglement entropy in some relevant geometries}

{\em Entanglement} or nonseparability refers to the existence of
quantum correlations between two sets of degrees of freedom of a
physical system that can be considered as subsystems. It is natural
that two (sub)systems in interaction are entangled an that they still
are entangled after their interaction has ceased. Particularly
interesting situations arise when two entangled systems become causally
disconnected because of an event horizon.

\subsection{Introduction: entanglement entropy}

The entanglement of two parts of a quantum system can be measured by
the von Neumann entropy. This is defined in terms of the density
matrices of either part. Let us consider, for later convenience, one
part as ``left'' or ``interior'' and another as ``right'' or
``exterior'', in a yet imprecise sense. Then, let us represent states
belonging to the left (or interior) with small letters and states
belonging to the right (or exterior) with capital letters. A basis for
the global states (left plus right) is
$\{|a\rangle\}\otimes\{|A\rangle\}$. By representing the ground state
in this basis as
\begin{equation}
|0\rangle=\sum_{aA} \psi_{aA}\, |a\rangle \otimes |A\rangle,
\end{equation}
we define a coefficient matrix $\psi_{aA}$.
Then we have two {\em different} density matrices:
\begin{equation}
\rho_{R}=
\frac{ {\psi}^\dagger{\psi}} {{\rm Tr}\,{\psi}^\dagger{\psi} }\,, \quad
\rho_{L}=
\frac{ {\psi}^*{\psi}^{{^T}}}{ {\rm Tr}\,{\psi}^*{\psi}^{{^T}} }\,.
\end{equation}
Correspondingly, we have two von Neumann entropies:
\begin{enumerate}
\item $S_{R}=-{\rm Tr}_A\left(\rho_{R}\ln \rho_{R}\right)$
\item $S_{L}=-{\rm Tr}_a\left(\rho_{L}\ln \rho_{L}\right)$
\end{enumerate}
Now it is important to recall the ``symmetry theorem'', which states that
both entropies are equal, $S_{R}= S_{L}$ (this can be
proved in several ways, the most popular one appealing to the {\em
Schmidt decomposition} of the entangled state.
The equality of entropies implies that they are associated
with properties shared by both parts, that is, with (quantum) correlations.

More generally, two non-interacting parts of a quantum system can be
originally in respective mixed states. After their interaction, which
we describe as an arbitrary unitary evolution of the composite system,
the initial density matrix $\rho_{L} \otimes \rho_{R}$ has evolved to 
$\rho'_{LR}$. It is easy to see that the partial traces 
$\rho'_{L}$ and $\rho'_{R}$, in general, have von Neumann entropies 
$S'_{L}$ and $S'_{R}$ such that $S'_{L} + S'_{R} \geq S_{L} + S_{R}$ 
\cite{Peres}. Of course, if the initial state is pure 
$S_{L} = S_{R} = 0.$

\subsection{Field theory half-space density matrix}

Let us now consider the quantum system to be a chain of coupled
oscillators. Moreover, we shall chiefly work in the continuum limit,
where the concepts and mathematical expressions are more transparent,
in spite of dealing with non-denumerable sets of degrees of freedom
(we will return to a discrete chain to describe the density matrix 
renormalization group algorithm).
The action for this model, namely, a one-dimensional scalar field, is
\begin{equation}
A[\varphi(x,t)] = \int\! dt\, dx \left({1\over 2}
\left[(\partial_t\varphi)^2 -
(\partial_x \varphi)^2 \right] - V(\varphi) \right),
\label{I}
\end{equation}
where $\varphi$ is the field.

Let us obtain a path integral representation for the density matrix
on the half-line of a system that is in its ground state 
\cite{Bomb,KabStr,CalWil}.
In the continuum limit, the half-line density matrix is a functional integral,
\begin{equation}
\r[\varphi_R(x),\varphi'_R(x)] =
\int\! D\varphi_L(x)\, \psi_0[\varphi_L(x),\varphi_R(x)]\,
\psi^*_0[\varphi_L(x),\varphi'_R(x)],
\label{DM0}
\end{equation}
where the subscripts refer to the left or right position of the
coordinates with respect to the boundary (the origin). Now, we must
express the ground-state wave-function as a path integral,
\begin{equation}
\psi_0[\varphi_L(x),\varphi_R(x)] = \int D\varphi(x,t)\, \exp\left(-A[\varphi(x,t)]\right),
\end{equation}
where $t\in (-\infty,0]$ and with boundary conditions 
$\varphi(x,0) =
\varphi_L(x)$ if $x<0$, and $\varphi(x,0) = \varphi_R(x)$ if 
$x>0$.
The conjugate wave function is given by the same path integral and
boundary conditions but with $t\in [0,\infty)$.  Substituting into
Eq.~(\ref{DM0}) and performing the integral over $\varphi_L(x)$, one
can express $\r(\varphi_R,\varphi'_R)$ as a path integral over
$\varphi(x,t)$, with $t\in (-\infty,\infty)$, and boundary conditions
$\varphi_R(x,0+) = \varphi'_R(x)$, $\varphi_R(x,0-) = \varphi_R(x)$.
In other words, $\r(\varphi_R,\varphi'_R)$ is represented by a single
path integral covering the entire plane with a cut along the positive
semiaxis, where the boundary conditions are imposed.

\begin{figure}
  \epsfxsize=5cm
\epsfbox{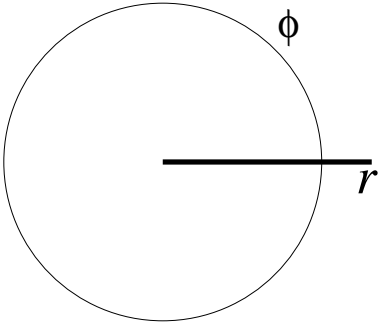}
\caption{Geometric setting for computing $\r(\varphi_R,\varphi'_R) = \langle
  \varphi'_R |\exp (-2\pi\,\H) |\varphi_R\rangle$, with angular
  Hamiltonian $\H$ and temperature $1/(2\pi)$.}
\label{circle}
\end{figure}

\subsection{Angular quantization and Rindler space}

In Euclidean two-dimensional field theory,
the generator of rotations in the $(x,t)$ plane is given by
\begin{equation}
\H =  \int dx\,(t\,T_{11}-x\,T_{00}),
\label{L}
\end{equation}
in terms of the components of the stress tensor computed from 
the action (\ref{I}). 
To simplify, one can evaluate it at $t=0$. 
Let us consider a free action ($V=0$). 
In the Schr\"odinger
representation, we should replace the momentum $\Pi = \partial_t\varphi$
with $\Pi(x) = i\,\delta/\delta \varphi(x)$.  
However, as in canonical quantization,
one rather uses the second-quantization method, which diagonalizes the
Hamiltonian by solving the classical equations of motion and
quantizing the corresponding normal modes.
Let us recall that, in canonical quantization,
if we disregard anharmonic terms,
the classical equations of motion in the continuum limit become
the Klein-Gordon field equation, giving rise to the usual
Fock space.  Not surprisingly, the eigenvalue equation for $\H$ leads
to the Klein-Gordon equation in polar coordinates in the $(x,t)$
plane,

The free field wave equation in polar coordinates,
\begin{equation}
(\Delta + m^2) \varphi = \left({1\over r}{\partial \over \partial
r}r{\partial \over \partial r} + {1\over r^2}{\partial^2 \over
\partial \phi^2}+ m^2\right) \varphi = 0,
\end{equation}
can be solved by separating the angular variable: it becomes a Bessel
differential equation in the $r$ coordinate with complex solutions
$I_{\pm i\,\ell}(m\,r)$, $\ell$ being the angular
frequency. We have a continuous spectrum, which becomes discrete on
introducing boundary conditions. One of them must be set at a short
distance from the origin, to act as an ultraviolet regulator
\cite{Bomb,KabStr,CalWil}, necessary in the continuum limit.

Therefore, the second-quantized field is (on the positive semiaxis
$t=0 \Leftrightarrow \phi = 0$, $x \equiv r$)
\begin{equation}
\varphi(x) = \int {d\ell\over 2\pi}\,\frac{b_\ell\,I_{i\,\ell}(m\,x) +
b_\ell^\dag\,I_{-i\,\ell}(m\,x)}{\sqrt{2\,\sinh(\pi\,\ell)}},
\end{equation}
where we have introduced annihilation and creation operators and
where the term that appears in the denominator is just for normalization,
to ensure that those operators satisfy
canonical conmutations relations.
There is an associated Fock space built by acting with $b_\ell^\dag$
on the ``vacuum state''. These states constitute the spectrum of
eigenstates of $\H$, which adopts the form
$\H = \int d\ell \,\ell\, b_\ell^\dag b_\ell$ (where the integral is
replaced with a sum for discrete $\ell$).

The type of quantization just exposed was first introduced in the 
context of quantization in curved space, in particular, in Rindler space.
Rindler space is just Minkowski space (therefore, not curved) 
in coordinates such that the time is
the proper time of a set of accelerated observers. Its interesting feature is 
the appearance of an event horizon, which implies that the ground state
(the Minkowski vacuum) is a mixed (thermal) state \cite{BD}.
The connection with black hole entropy and Hawking radiation
is explained in the next section.

It is pertinent to note that 
the functions $I_{\pm i\,\ell}(m\,x)$ have wave-lengths that increase with
$x$.  It is illustrative to represent a real ``angular wave'',
$$K_{i\,\ell}(m\,x) = \frac{i\,\pi}{2\,\sinh(\pi\,\ell)}\,
[I_{i\,\ell}(m\,x)-I_{-i\,\ell}(m\,x)].$$
This solution is oscillatory for $x < \ell/m$,
with a wavelength proportional to $x$, and decays
exponentially for $x > \ell/m$ (Fig.~1). The fact that the wave length
vanishes at $x=0$ is to be expected from the Rindler space viewpoint, because 
it corresponds to the horizon.

\begin{figure}
  \epsfxsize=9cm
\epsfbox{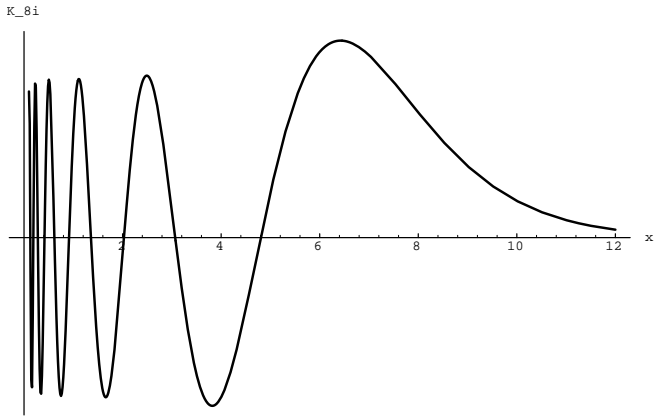}
\caption{Real radial wave $K_{8\,i}(x)$.}
\label{wave}
\end{figure}

\subsection{Black hole entropy}

Let us consider the Schwarzschild geometry in the 
Kruskal-Szekeres coordinates $u,v$, defined by
$$u\,v =16M^2 \left(\frac{r}{2M}-1\right) \exp\left(\frac{r}{2M}-1\right), 
\quad \frac{u}{v} = \exp\frac{t}{2M}.$$
The horizon is given by $u=0$ or $v=0$, like in the Rindler geometry of the 
previous section. Then, if we further define
$$Z = u + v,\quad T = u- v,$$ these coordinates behave like the
ordinary Minkowskian coordinates.  Moreover, for small $u$ or $v$ 
(or the large $M$ limit), the
curvature can be neglected and the geometry becomes locally the one of
Rindler space. Therefore, we could define radial coordinates $Z = \r\,
\cosh \tau,$ $T = \r \,\sinh \tau,$ and perform a radial quantization
like we did in the preceding section.

Once established that the geometry near the black-hole horizon is locally
the Rindler geometry of the preceding section, 
we can readily transfer the form of the density
matrix of a scalar field therein, 
where we now ignore (trace over) the degrees of 
freedom inside the horizon. Hence, we can define a
von Neumann entropy associated with this density matrix. 
Furthermore, in so doing, we can appreciate that the concept of black-hole
entropy takes a new meaning: in addition to being of quantum origin, 
this entropy is related with 
shared properties between interior and exterior, namely, with the horizon.
In addition, the radial vacuum is a thermal state with respect to the original 
Schwarzshild coordinates, giving rise to Hawking radiation \cite{BD}.

\section{Quantum information and RG transformations}  

\subsection{Information theory and maximum entropy principle}

The entropy concept appeared in Thermodynamics but only took a truly
fundamental meaning with the advent of information theory. In this
theory, entropy is just missing information, while information itself
is often called {\em negentropy}. To recall basic definitions, the
information attached to an event that occurs with probability $p_n$ is
$I_n = -\ln p_n.$ Hence, the average information (per event) of a
source of events is $S(\{ p_n \}) = \sum_n p_n I_n = - \sum_n p_n \ln
p_n.$

The previous definitions, given by Shannon in his theory of
communication, seem unrelated to thermodynamic entropy as a property
of a physical system. However, according to the foundations of
Statistical Mechanics on Probability Theory (the Gibbs concept of
ensembles), a clear relation can be established, as done by Jaynes
\cite{Jay}. Jaynes made connection with the Bayesian philosophy of
probability theory, in which the concept of ``a priory'' knowledge is
crucial.  Indeed, although the exact microscopic state of a system
with many degrees of freedom may be unknown, one has some ``a priory''
knowledge given by the known macroscopic variables. This is a particular
case of Jaynes' adaptation of the Bayesian probability theory, wich
postulates that the best probability distribution to be attributed to
a stochastic event is such that it incorporates only the ``a priory''
knowledge about the event and nothing else.  This postulate amounts to
Jaynes' {\em maximum entropy} principle: given some constraints, one
must find the maximum entropy probability distribution (density
matrix, in the quantum case) compatible with those constraints,
usually, by implementing them via Lagrange multipliers.  In
particular, more constraints mean less missing information and lead to
less entropy.

\subsection{Quantum information}

The concepts of Shannon's classical theory of communication have
quantum analogues \cite{Peres,Schu}. Nevertheless, the quantum theory of
communication is richer (and less intuitive!).  Indeed, the key new
notion in the quantum theory is entanglement (already described in the
foregoing): if a state (an event) is entangled with the environment,
we have the type of purely quantum phenomena to which the EPR paradox
is associated.

Schumacher posed the problem of communication of an entangled state
\cite{Schu}. (The technical name is {\em transposition}, since the
copy of a quantum state is not possible: no-cloning theorem
\cite{Peres}.) His conclusion is that the von Neumann entropy of the
state is the quantity that determines the {\em fidelity} of the
transposition: it is possible to transpose the state with near-perfect
fidelity if the signal can carry at least that information.

The fidelity is simply defined as the overlap between normalized
states, namely, $\left|\langle \psi | \psi'\rangle\right|^2$. It is
directly related with the natural distance in the space of rays, that
is, the angle between rays. The geometrical meaning of this distance
is best perceived by considering the complex projective 
space of rays, where it is called the Fubini-Study metric
\cite{Gib}.%
\footnote{It is very interesting to realize that the distance between
the distribution probabilities defined by the outcomes of possible
measurements of observables that distinguish the states $\psi$ and
$\psi'$ coincides with the statistical distance in the classical sense
of Fisher \cite{Woo,BraCa}.}  Schumacher maximization of the fidelity
uses the Schmidt decomposition of the entangled state \cite{Schu}.

\subsection{Renormalization group and information theory}

The problem of transforming one quantum state into another while
preserving its fundamental features is reminiscent of an operation
performed with the quantum renormalization group (in this connection, 
see Ref.\ \cite{Pres}). This operation does
not intend to reach near-perfect fidelity but it is desirable that it
reach as much fidelity as possible. We shall see that a particular
formulation of the renormalization group, namely, the density matrix
renormalization group, comes close to applying the concepts of quantum
information theory, and in a similar way to Rindler space quantization
\cite{I}. In fact, the construction of the density matrix
renormalization group algorithm is based on the Schmidt decomposition,
in parallel to Schumacher fidelity maximization.

Let us consider an entangled quantum state $|\psi\rangle$ for which we
seek an optimal reduced representation $|{\tilde\psi}\rangle$ (a sort of 
{\em quantum coding}). That is, we must find a projection 
\begin{equation}
|\psi\rangle = \sum_{aA} \psi_{aA} \,|a\rangle \otimes |A\rangle
\ra |{\tilde\psi}\rangle = \sum_{iA} \psi_{iA} \,|i\rangle \otimes |A\rangle
\end{equation}
to a subspace spanned by reduced basis $\{i\}$
such that the distance 
$S = | |{\tilde\psi}\rangle - |\psi\rangle |^2$ is minimized. This is
the problem that S.\ White met in looking for an improved ``real
space'' quantum renormalization group algorithm: In a renormalization
group step, one must reduce the number of block states, but in a way
that accounts for the influence of the rest of the system (that is,
with no introduction of arbitrary boundary conditions)
\cite{White}. This amounts to consider the entanglement of the block
and to formulate the above described problem. White found the solution
in terms of the {\em singular value decomposition}, a well-known
numerical algorithm, equivalent to the Schmidt decomposition of the
entangled state and, hence, leading to discarding the smallest
eigenvalues of the block density matrix; in other words, $|i\rangle$
are the eigenstates of $\rho$ with largest eigenvalues.
The proper formulation of the density matrix
renormalization group in terms of quantum information concepts 
has been provided recently \cite{OsbNiel}.

\subsubsection{Density matrix renormalization group algorithm}    

Method for a $1D$ quantum system on a chain (e.g., a chain of
oscillators):

  \begin{enumerate}
\item Select a sufficiently small ``soluble'' block $[0,L]$:
\hspace{1cm}\includegraphics[width=4cm]{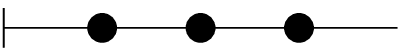}
\item Reflect the block on the origin: 
\hspace{1cm}\includegraphics[width=7cm]{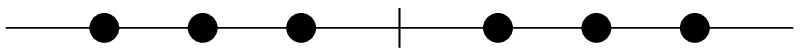}
\item Compute the ground state.
\item Compute the density matrix of the block $[0,L]$.
\item Discard eigenstates with smallest eigenvalues.
\item Add one site next to the origin.
\item Go to 2.
  \end{enumerate}

One has to adjust this procedure in such a way that the iteration
keeps the Hilbert space size approximately constant. The procedure can
be performed algebraically for a chain of coupled {\em harmonic}
oscillators \cite{I}. Otherwise, it has to be performed numerically.
In the continuum limit, the connection with the scalar field theory
half-space density matrix previously calculated is clear.  In
particular, the property of ``angular waves'' of having vanishing
wave-length at $x=0$ pointed out above provides an explanation of the
form in which the density matrix renormalization group solves the
boundary condition problem, that is, because of the concentration of
quantum states near the boundary.  To be precise, using the
eigenfunctions of $\H$ instead of free waves, we have a basis in which
the region close to $x=0$ (the boundary point) is more accurately
represented than the region far from it when we cut off the higher
$\ell$ eigenfunctions.

\section{Generalizations and applications}

\subsection{Geometric entropy}

We have seen that the half line density matrix of a field theory has a
geometric interpretation in Rindler space. Furthermore, the entropy 
of black holes can be understood as a generalization to a more complicated 
geometry. We may wonder if further generalizations are possible. 

In this connection, we recall the notion of ``geometric entropy'',
introduced by C.\ Callan and F.\ Wilczek \cite{CalWil}, as the entropy
``associated with a pure [global] state and a geometrical region by
forming the pure state density matrix, tracing over the field
variables inside the region to create an `impure' density matrix''.
Of course, their motivation was the earlier suggestion that 
back-hole entropy is of quantum-mechanical entanglement origin \cite{Bomb}.
They proposed a generalization to different topologies but, actually, they
only computed the Rindler space case, discussing the divergence 
of the entropy at the horizon \cite{CalWil} (the UV 
divergence of this type of entropy had been discussed in general in Ref.\ 
\cite{Bomb}). 

A different notion of geometric entropy can be deduced by purely
geometrical means from the presence of horizons, namely, as
associated with a spacetime topology that does not admit a trivial
Hamiltonian foliation \cite{HH}. This type of topology prevents 
unitary evolution and leads to mixed states.

In fact, it is only the second type of entropy that embodies the
famous ``one-quarter area law'' for black holes, due to its origin in
purely relativistic concepts (this was demonstrated for an earlier
relativistic notion of entropy in Ref.\ \cite{GH}).  On the contrary,
the quantum notion of geometric entropy for a field theory involves UV
divergences and needs renormalization before a comparison with 
the relativistic notion can be made
(see the discussion in Ref.\ \cite{Bek}).

\subsection{Application to cosmology}

The generalization of the concept of black-hole entropy to the de
Sitter space and, hence, to cosmology is relatively old \cite{GH2}.
Of course, the concept of entropy in its traditional thermodynamical
sense has been crucial in explaining the dynamics of {\em inflation}
(now a standard paradigm), namely, in accounting for the {\em
reheating} process (entropy generation). However, the relations
between the traditional view and the one associated to quantum
entanglement may lead to further insight, when they are properly 
formulated in the context of quantum information theory.

In particular, we can regard the generation of entropy and
fluctuations in de Sitter spacetime as a fundamentally quantum process
leading to the celebrated Harrison-Zeldovich {\em scale invariant}
spectrum of Gaussian fluctuations. If we consider the initial state
for inflation as a pure quantum state (that some theory of quantum
gravity will hopefully characterize some day), the de Sitter space
horizon induces {\em decoherence} of the modes which cross it,
irrespective of the actual inflaton dynamics and, therefore, of the
details of the reheating process. This decoherence consists of a
randomization of the phases of the present quantum fields and
naturally produces {\em thermal} Gaussian fluctuations. Moreover, the
symmetry of de Sitter space implies that each mode has the same
physical size as it crosses the horizon, leading to the
Harrison-Zeldovich power spectrum:
$$P(k) \equiv \left| \left( \frac{\delta \r}{\r_0} \right)_{\!\!k}
\right|^2 \propto k\,.$$ Actually, this spectrum is given by the
equipartition theorem at the corresponding {\em Hawking temperature}
$T = H/(2\pi)$ ($H = \Lambda/3$) \cite{GH2}.

In this reasoning, one could also consider the initial state to 
be mixed (e.g., a thermal state) and apply the argument for entropy 
growth exposed in subsection 1.1.

One may wonder if a density-matrix type renormalization group could help 
with the dynamics. As long as the fluctuations are Gaussian, we are in a 
similar situation to the case of harmonic oscillators commented above, which 
makes any renormalization group superfluous. However, as is well known, 
the gravitational instability produces {\em non-linear} evolution and leads 
to phase correlations. Therefore, the ideas presented here may be useful in 
setting up a renormalization group for the study of 
{\em non-Gaussian} fluctuations and its non-linear evolution. 
In particular, the Wilson or {\em exact} renormalization group 
irreversibility properties and can be connected with other methods 
of analysis of the non-linear evolution 
(for preliminary steps in that direction, see Ref.\ \cite{I2}).
\vskip 1cm

{\bf\large Acknowlegments}
\vskip 0.5cm

I am grateful to Juan Poyatos for bringing some relevant references 
to my attention.

\end{document}